\documentclass[aps,prd,amssymb,showpacs,floatfix,twocolumn,superscriptaddress,nofootinbib]{revtex4-1}
\usepackage{amsmath,amsfonts,graphics,epsfig}
\begin{document}
\title{The radius of the $\rho$ meson determined from its decay constant}
\author{A.F.~Krutov} \email{krutov@samsu.ru},
\affiliation{Samara State University, 443011 Samara, Russia}
\author{R.G.~Polezhaev} \email{polezaev@list.ru},
\affiliation{Samara State University, 443011 Samara, Russia}
\author{V.E.~Troitsky} \email{troitsky@theory.sinp.msu.ru}
\affiliation{D.V.~Skobeltsyn Institute of Nuclear Physics,\\
M. V. Lomonosov Moscow State University, Moscow 119991, Russia}
\date{\today}
\begin{abstract}
We present a unified model describing electroweak properties of
the $\pi $ and $\rho $ mesons. Using a general method of
relativistic parametrization of matrix elements of local
operators, adjusted for the nondiagonal in total angular momentum
case, we calculate the $\rho $-meson lepton--decay constant
$f_{\rho}$ using the same parameters of free constituent quarks
that have ensured exclusively good results for the $\pi$ meson
previously. The only free parameter, characterising quark
interactions, which include additional spin-spin contribution and
hence differ from the $\pi$-meson case, is fixed by matching the
decay constant to its experimental value. The mean square charge
radius is calculated, $\langle r^2_{\rho}\rangle  = \left( 0.56
\pm 0.04 \right)$~fm$^{2}$. This result verifies, for the
$\rho$-meson case, the conjecture of equality between
electromagnetic and strong radii of hadrons tested previously for
proton, $\pi$ and $K$ mesons.

\end{abstract}

\pacs{11.10Jj, 12.39Ki, 13.40Gp, 14.40Aq}
\maketitle

\section{Introduction}

The construction of effective quantitative
methods to calculate electroweak properties of hadrons is an important
direction in composite-particle physics. In numerous respects, $\pi$ and
$\rho$ mesons, consisting of light quarks, are the simplest bound states to
study.

While the pion properties are well known from experiments, the
situation is quite different in the case of $\rho $ meson. Its
lifetime is very short, $\sim 4.5\times10^{-24}$~s, so direct
measurements of its electroweak properties (e.g.\ electromagnetic
form factors and static moments) are nearly impossible. The fact
that experimental data on the $\rho $ meson are scarce brings one
to the natural difficulties while choosing the most adequate
approaches to understanding its structure, estimating
approximations behind chosen methods and determination of
parameters of models. However, knowledge of electroweak properties
of the $\rho$ meson, which is one of the simplest hadrons, is a
necessary ingredient of understanding physical properties of
composite strongly interacting systems. In particular, as it is
discussed below, nontrivial conjectures about meson properties
exist whose testing requires knowing $\rho$-meson properties not
accessible experimentally. That is why, during last years, a
number of papers, e.g.\ Refs.~\cite{BhM08, RoC11, RoB11, AlO09,
CaB15, LoM00, HeK07, LeM08, ChA15, OwK15, GrR07, MeS02, BiG14}, on
the $\rho $ meson did appear although their results cannot be
directly compared to measurements.

The problem was attacked by different approaches, including
well-known methods in the frameworks of QCD and
based on the Dyson-Schwinger equation \cite{BhM08, RoC11, RoB11}, QCD sum
rules \cite{AlO09}, Nambu--Jona-Lasino model \cite{CaB15}, the hybrid
model \cite{LoM00}, lattice QCD approach, that is now the only one giving
the direct calulation of form factors using QCD, \cite{HeK07, LeM08,
ChA15, OwK15}, and also the holographic approach \cite{GrR07}, the
light--front formalism and the Feinman triangle diagram \cite{MeS02}, the
Covariant Spectator Theory \cite{BiG14}.

It is important to point out different relativistic formulations
of constituent quark models that are based on the classical paper
by P.~Dirac \cite{Dir49} (so-called Relativistic Hamiltonian
Dynamics or Relativistic Quantum Mechanics (RQM), see, e.g. the
reviews \cite{LeS78,KeP91,Coe92,KrT09}). This approach was used to
describe composite quark systems in Refs.
\cite{Kei94,CaG95,KrT02,KrT03,BaC02,ChC04,HeJ04,ChC07,BiS14,MeS15}.
Various calculations use essentially different bases,
approximations and parameters give essentially different
predictions for electroweak properties of $\rho $-meson. For
instance, even within the frameworks of light-front dynamics, the
results vary because of different treatment of the so-called
angular condition.

To calculate the electromagnetic characteristics of $\rho $-meson
in the present paper we use our version of the instant form (IF)
of RQM. This model is described in detail in Refs.
\cite{KrT01,KrT02,KrT03,KrT09prc,KrT09}. The main differences
between our version and the conventional IF RQM are, first, the
construction of the electroweak current matrix elements which
based on the analogue of the Wigner-Eckart theorem for the
Poincar\'e group and, second, the interpretation of the
corresponding reduced matrix elements, that is form factors
\cite{KrT05,Edm55} as generalized functions.  We include the
interaction in the composite system by adding the interaction
operator to the operator of the mass of the free constituent
system by analogy with conventional IF RQM. Note that it
is possible to include the interaction in our approach by the
solutions of the Muskhelishvili-Omnes type equations \cite{ShT69}.
These solutions represent wave functions of constituent quarks.


It is important to notice that the approach we use differs from
the IF \textit{per se} but it was rather fruitfully complemented
by the so-called Modified Impulse Approximation (MIA), see Ref.
\cite{KrT02}. MIA is constructed by making use of a
dispersion--relation approach in terms of the reduced matrix
elements (form factors) and removes certain (often quoted)
disadvantages of IF. In particular, in our model, the
Lorentz-covariance condition and the current conservation law are
satisfied automatically.

This approach was surprisingly predictive in description of the
pion form \cite{KrT01,KrT09prc}: subsequently obtained
experimental data (see Ref.\cite{Hub08} and references therein)
for the range of momentum transfers, $Q^{2}$, larger by an order
of magnitude, coincide precisely with prediction of Ref.
\cite{KrT01} without any further tuning of parameters. Moreover,
and remarkably, this is the only available low-energy model which
reproduces the correct QCD asymptotics of the pion form factor
\cite{FaJ79,EfR80,LeB79} without tuning of additional parameters.
It worths to make a special emphasis on the fact that our approach
gives not only the correct QCD power-law in asymptotic decreasing
of pion form factor at large momentum transfers \cite{KrT98}, but
the asymptotics QCD prefactor as well \cite{TrT13,TrT15}. The
soft/hard transition was governed by the switching the
constituent-quark mass $M(Q^{2})$ off. The physical reasons for
such a coincidence require, certainly, an additional
investigation.

The model reproduces the standard nonrelativistic limit correctly.
The method permits analytic continuation of the pion
electromagnetic form factor from the spacelike region to the
complex plane of momentum transfers and gives a good description
of the pion form factor in the timelike region \cite{KrN13}.

The main purpose of the present paper is the construction, in the
frameworks of IF of RQM, of the unified model describing $\pi$ and $\rho$
mesons, and calculation of the mean square charge radius
$ \langle r^2_{\rho}\rangle $
of the $\rho $ meson.

We take advantage of our successful model of pion which fixes the
light-quark parameters. There remains only one unknown parameter,
the wave function parameter $b$, describing the effective
interaction in the two-quark system (because of the spin-spin
interaction, it is different for $\pi$ and $\rho$ mesons). We
obtain an explicit expression for the $\rho $-meson decay constant
that depends, now, only on $b$, $f_{\rho}(b)$. Then we fix the
value of $b$ from the experimental value of $f_{\rho}$. Finally,
with this fixed $b$, we  obtain
the mean square radius $\langle r^2_\rho\rangle$.

The rest of the paper is organized as follows. In Sec.~\ref{sec:f_rho}, we
obtain the analytic expression of the lepton decay constant for vector
mesons using the instant form of relativistic quantum mechanics.  For this
purpose we use the general method of relativistic invariant
parametrization of matrix elements of local operators nondiagonal in total
angular momentum \cite{KrP15}. In Sec.~\ref{sec:param}, the actual formulae
that are used for calculations are given and the process of parameter
fixing is described. The results of the calculation of the $\rho$-meson
charge radius are presented. Sec.~\ref{sec:discussion} discusses the
results while Sec.~\ref{sec:concl} contains our
conclusions.

\section{The lepton decay constant of the $\rho$ meson in the instant form
of RQM}
\label{sec:f_rho}

The lepton decay constant of a vector meson, $f_{c}$, is defined
by the following matrix element of the electroweak current (see, e.g.
\cite{Jau03}):
\begin{equation}\label{const11}
\langle0|j^c_{\mu}(0)|\vec{P_{c}},m_{c}\rangle =
i\,\sqrt{2}\,f_{c}\,\xi_{\mu}(m_c)\frac{1}{(2\pi)^{3/2}}\;,
\end{equation}
where $\vec{P_{c}}$ is the meson three-momentum, $m_{c} = -1,0,1$ is
the spin projection, $\xi_{\mu}(m_c)$ is the polarization vector
that in the Breit frame (BF) has the form
\begin{equation}\label{br}
\xi_{\mu}(\pm1) = \frac{1}{\sqrt{2}}(0,\mp1,-i,0)\;, \quad
\xi_{\mu}(0)=(0,0,0,1)\;.
\end{equation}

Now our aim is to obtain the lepton decay constant in terms of the
wave function of the meson considered as two-quark composite
system in the frameworks of the variant of IF of RQM that was
developed by the authors (see, e.g. \cite{KrT02, KrT03,KrT09,
KrP15}). To solve this problem, we must, first, decompose the
matrix element in (\ref{const11}) in some basis that define the
representation of the wave function, and, second, separate the
invariant part in the decomposition and therefore parametrize the
matrix element \cite{KrP15}.  Let us note that the form
(\ref{const11}) with segregated invariant part $f_c$ presents a
particular case of parametrization. Nevertheless we will use a
more universal procedure proposed in \cite{ChS63} for the current
matrix elements which are diagonal in total moment and then
developed for the case of composite systems \cite{ShT69,KoT72} and
for decay processes \cite{KrP15}.

In RQM the state vector of a two-particle system belongs to the direct
product of two one-particle Gilbert space. So we can describe it in the two
bases.

1. The basis of individual spins and momenta of two particles:
\begin{eqnarray}
\label{dv}
&& \hspace{-2mm}|\vec{p_1},m_1;\vec{p_2},m_2\rangle = |\vec{p_1},m_1\rangle
\otimes{|\vec{p_2},m_2}\rangle\;,\nonumber\\[2mm]&&\langle\vec p, m|\vec
p\;',m'\rangle = 2\,p_0\,\delta(\vec p - \vec
p\;')\,\delta_{m\,m'}\;,
\end{eqnarray}
where $\vec p_{1,2}$ are three-momenta of particles,
$m_{1,2}$ are spin projections, $p_0^2 - \vec p\,^2 = M^2$, $M$ is the
mass. In what follows the masses of particles are supposed to be equal.

2. The basis where the center-of-mass motion is separated:
\begin{equation}
\label{vec}
|\vec{P},\sqrt{s},J,L,S,m\rangle\;,
\end{equation}
where $\vec{P} = \vec{p_{1}}+{\vec{p_{2}}}$,  $\sqrt{s}$ is the invariant
mass of a system of two free particles, $P^2 = s,\;$ $L$ is the orbital
momentum in the center-of-mass system (CMS) and $S$ is the total spin in
CMS.

The state vector (\ref{vec}) is normalized as
\begin{eqnarray}\label{1}
&& \hspace{-2mm}\langle\vec{P},\sqrt{s},J,L,S,m|\vec{P}\;',\sqrt{s'},J',L',S',m'\rangle
=
\nonumber\\[2mm]&& N\,\delta(\vec{P}-\vec{P}\;')\,
\delta(\sqrt{s}-\sqrt{s'})\,\delta_{JJ'}\,\delta_{LL'}
\,\delta_{SS'}\,\delta_{mm'}\;,
\end{eqnarray}
where $N$ is the normalization constant, whose explicit form is irrelevant
in what follows.

The bases (\ref{dv}) and (\ref{vec}) are related by the Clebsch-Gordan
decomposition for the Poincar\'e group
(see, e.g., \cite{KrT09}):
\begin{eqnarray}\label{Clebsh111}
&& \hspace{-2mm} |\vec{P},\sqrt{s},J,L,S,m\rangle =
\sum_{m_{1},m_{2}}\int\,\frac{d^3{\vec{p_1}}}{2p_{10}}\frac{d^3{\vec{p_2}}}{2p_{20}}
|\vec{p_1},m_{1};\vec{p_2},m_{2}\rangle
\nonumber\\[1mm]&& \times\langle\vec{p_1},m_{1};\vec{p_2},m_{2}|\vec{P},\sqrt{s},J,L,S,m\rangle\;,
\end{eqnarray}
where
\begin{eqnarray}\label{cli}
&&\hspace{-2mm}
\langle\vec{p_1},m_{1};\vec{p_2},m_{2}|\vec{P},\sqrt{s},J,L,S,m\rangle
= \frac{2\sqrt{s}}{\sqrt{\lambda(s,M^2,M^2)}}
\nonumber\\
[2mm]&& \hspace{-2mm} \times 2P_{0}\delta(P-p_{1}-p_{2}) \,
\sum_{\tilde{m_{1}},\tilde{m_{2}}}D^{1/2}_{m_1\tilde{m}_1}(p_{1},P)
D^{1/2}_{m_2\tilde{m}_2}(p_{2},P)\, \nonumber\\
[2mm]&& \hspace{-2mm}\times\langle
\frac{1}{2}\tilde{m_{1}}\frac{1}{2} \tilde{m_{2}}|S m_{S}\rangle
Y_{L m_{L}}(\vartheta,\varphi)\langle S L m_{S} m_{L}|J
m\rangle\;,
\end{eqnarray}
$\lambda(a,b,c) = a^2+b^2+c^2-2(ab+ac+bc)$, $D^{1/2}$ are the
matrices of three dimensional rotations (Wigner
$D$-functions),$\;Y_{L m_{L}}$ are the spherical harmonics.

In Eq.~(\ref{cli}) we expand in spherical harmonics and sum over
angular momenta in the CMS and then pass to an arbitrary reference
frame using Wigner $D$-functions.

We calculate the lepton decay constant in Eq.~(\ref{const11})
in the case of the four-fermion interaction. So, we consider  the matrix
element of the electroweak current of decay $j\,^0_{\mu}$ of system of two
free fermions using the basis (\ref{vec}):
\begin{equation}\label{nodia11}
\langle 0|j\,^0_{\mu}(0)|\vec{P},\sqrt{s},J,L,S,m\rangle\;.
\end{equation}

Keeping in mind the $\rho$ meson, we consider, in what follows, only
vector mesons with  zero orbital moment of quark relative movement.
So, we put $J = S = 1$, $L=0$ and omit the corresponding variables
in the state vectors of the basis
(\ref{vec}).

Let us consider the relativistic invariant parametrization of the matrix
element
(\ref{nodia11}) following \cite{KrP15, ChS63}.
We perform a Lorentz transformation from the initial (laboratory)
frame to the BF,
$$
\vec{\tilde{j\,^0}} = \vec{j\,^0} +
\frac{\vec{w}(\vec{j\,^0}\,\vec{w})}{1+w_{0}} -
\vec{w}\,j\,^0_{0}\;,
$$
\begin{equation}\label{lor}
\tilde{j}\,^0_0 = j\,^0_{0}\,w_{0} - (\vec{j\,^0}\,\vec{w}) =
j\,^0_{\mu}\,w^{\mu}\;,
\end{equation}
where $w_{\mu}={P_{\mu}}/{\sqrt{s}}$ is the four-velocity corresponding to
this transformation,
$\tilde{j\,^0_\mu}$ is the four-vector of the electroweak current operator
in the BF.

In the case of the matrix element (\ref{nodia11}), the BF coincides with
the CMS where $\vec{\tilde{P}} = 0$, so the relation between matrix
elements in these systems is of a simple form (compare with Eq.~(21) in
\cite{KrP15}):
\begin{equation}\label{diagonal1111}
\langle 0|j\,_{\mu}^{0}(0)|\vec{P},\sqrt{s},m\rangle = \langle
0|\tilde{j}\,_{\mu}^{0}(0)|\sqrt{s},{m}\rangle\;,
\end{equation}

Eq.~(\ref{lor}) implies that the zero component of the operator,
$\tilde{j}\,^0_0$, in the BF has a scalar operator structure  or the
structure of a zero-rank spherical tensor operator. We must describe the
current operator in terms of tensor operators because we use the
Wigner-Eckart
theorem in what follows
 \cite{KrT05}.

Using the notations of \cite{KrP15} let us write:
$$
\langle 0|j\,_{0}^{0}(0)|\vec{P},\sqrt{s},m\rangle = \langle
0|\tilde{j}\,^0_0(0)|\sqrt{s},{m}\rangle =
$$
\begin{equation}\label{breitsfer11}
\frac{1}{\sqrt{4\pi}} \langle 0|\,C\,^{0}_{0}(s)|\,{m}\rangle\;,
\end{equation}
where $C\,^{0}_{0}$ is a spherical tensor operator of zero rank.

Performing the Wigner-Eckart decomposition of this spherical tensor
operator we obtain that its matrix element is zero:
\begin{equation}\label{breitsfer411}
\langle 0|\,C\,^{0}_{0}(s)|\,{m}\rangle = \langle 1\,{m}\, 0\,
0|\,0\,0\rangle\, G\,^{0,0}_{0,1}(s) = 0\;.
\end{equation}

We now consider the three-dimensional part of the current
operator matrix element (\ref{diagonal1111}):
\begin{equation}\label{breit111}
\langle 0|j\,^0_r(0)|\vec{{P}},\sqrt{s},m\rangle, \quad r =
1,2,3\;.
\end{equation}
The zero component of the current operator in BF (\ref{diagonal1111})
is treated as a rank-zero tensor operator.
We can describe the three-dimensional part of the operator in terms of a
rank-one tensor operator. To this end, it suffices to pass to the canonical
basis, i.e. to pass from Cartesian coordinates to the basis of
spherical harmonics \cite{Edm55}:
\begin{equation}\label{B}
\tilde{j}\,^0_r(0) = a_{r t}\tilde{j}\,^{01}_t(0), \quad t =
-1,0,1\;,
\end{equation}
\begin{equation}\label{C}
a_{r t}=\sqrt{\frac{2\pi}{3}}\left(
\begin{tabular}{ccc}
-1 & 0& 1\\
i & 0& i\\
0 & ${\sqrt{2}}$& 0\\
\end{tabular}\right)\;,
\end{equation}
where $\tilde{j}\,^{01}_t(0)$ are the components of the rank-one spherical
tensor operator.

Using the notations of \cite{KrP15} again, we can write:
$$
\langle 0|j^{01}_t(0)|\vec{P},\sqrt{s},m\rangle = \langle
0|\tilde{j}\,^{01}_t(0)|\sqrt{s},{m}\rangle =
$$
\begin{equation}\label{breitsfer1}
\frac{1}{\sqrt{4\pi}}\langle 0|B^{1,0}_{t,0}(s)|{m}\rangle\;,
\end{equation}
where $B^{1,0}_{t,0}$ is a spherical tensor operator transforming under
rotations upon the representation $D^{1}$.

Let us use now the Wigner-Eckart theorem
\begin{equation}\label{breitsfer41}
\langle 0|B^{1,0}_{t,0}(s)|\,m\rangle = \langle 1\,m\, 1\,
t|\,0\,0\rangle G^{1,0,1}_{0,1}(s)\;,
\end{equation}
where $G^{1,0,1}_{0,1}(s)$ is the set of reduced matrix elements, i.e.
scalar functions or the free form factors.

So, we obtain the parametrization of the three-dimensional part of the
current matrix element
(\ref{breit111}):
\begin{equation}\label{Atok11}
\langle 0|j\,^{01}_t(0)|\vec{P},\sqrt{s},m\rangle =
 \langle 1\,m\, 1\, t |\,0\,0\rangle\,\frac{1}{\sqrt{4\pi}}\,
G^{1,0,1}_{0,1}(s)\;.
\end{equation}

To obtain the explicit form of free form factors in (\ref{Atok11}), we
decompose the matrix element
(\ref{nodia11}) for previously fixed quantum numbers in the basis
(\ref{dv}),
\begin{eqnarray}\label{Clebsh1}
&&\hspace{-2mm}
\langle 0|j\,^0_\mu(0)|\vec{P},\sqrt{s},m\rangle =
\sum_{m_{1},m_{2}}\int\,\frac{d^3\,\vec{p_1}}{2p_{10}}
\frac{d^3\,\vec{p_2}}{2p_{20}}\,
\nonumber\\
[2mm]&& \hspace{-2mm}\times\,\langle\,
0|j^{0}_\mu(0)|\vec{p_1},m_{1};\vec{p_2},m_{2}\rangle\nonumber\\
[2mm]&& \times\,
\langle\vec{p_1},m_{1};\vec{p_2},m_{2}|\vec{P}\;,\sqrt{s},m\rangle\;,
\end{eqnarray}
where
$\langle\vec{p_1},m_{1};\vec{p_2},m_{2}|\vec{P},\sqrt{s},m\rangle$
is defined by the relation (\ref{cli}).

We use the standard expression (see, e.g. \cite{Jau03}) for the
current matrix element in the case of decay in the basis of individual
spins and momenta:
$$
\langle 0|j^{0\mu}(0)|\vec{p_1},m_{1};\vec{p_2},m_{2}\rangle =
$$
\begin{equation}\label{vf1}
\bar v(\vec{p_2},m_2)\gamma^{\mu}(1+\gamma^5)u(\vec{p_1},m_1) \;,
\end{equation}
$\gamma^{\mu}=(\gamma^{0},\vec{\gamma})$ are the Dirac matrices; $\bar
v(\vec{p_2},m_2)$ and $u(\vec{p_1},m_1)$ are the Dirac spinors.

Using Eqs.\ (\ref{breitsfer11}), (\ref{breitsfer411}),
(\ref{B}), (\ref{Atok11}), and integrating in
Eq.~(\ref{Clebsh1})  in BF, we obtain the explicit form of the free form
factors,
$$
G^{1,0,1}_{0,1}(s) =
$$
\begin{equation}
-\frac{3\sqrt{3}(\sqrt{s}+2M)}{16\sqrt{s}\pi^2}
\left(\frac{7s+12M\sqrt{s}+8M^2}{6s+12M\sqrt{s}+12M^2}\right)\;.
\label{freeff}
\end{equation}

Let us return now to Eq. (\ref{const11}) and decompose its l.h.s.\ in the
basis (\ref{vec}):
$$
\langle 0\,|j\,^c_\mu (0)|\vec P_c\,,m_{c}\rangle =
\sum_{m}\int\,\frac{d^3\vec P}{N} d\sqrt{s}\langle 0\,|j\,^c_\mu
(0)|\vec{P},\sqrt{s},m\rangle
$$
\begin{equation}\label{qq}
\times\langle\vec{P},\sqrt{s},m\,|\vec{P_{c}},m_{c}\rangle\;,
\end{equation}
where $\langle\vec P\,,\sqrt{s}\,,m|\vec P_c\,,m_{c}\rangle $ is the wave
function in the RQM represented in the basis (\ref{vec}),
\begin{equation}\label{vf}
\langle\vec P\,,\sqrt{s}\,,m|\vec P_c\,,m_c\rangle =
N_c\,\delta(\vec P - \vec P_c)\,\delta_{m\,m_c}\,\varphi(s)\;,
\end{equation}
where
$$
N_c = \sqrt{2\,P_{c0}}\,\sqrt{\frac{N_{CG}}{4\,k}}\;,\quad N_{CG}
= \frac{(2\,P_0)^2}{8\,k\,\sqrt{s}}\;.
$$
$$
k = \frac{1}{2}\sqrt{s - 4\,M^2}\;,
$$
\begin{equation}\label{1}
\varphi(s)= \sqrt[4]{s}\,k\,\psi({k})\;
\end{equation}
and $\psi({k})$ satisfies the normalization condition:
\begin{equation}\label{q1}
\int\psi^2(k)\,k^{2}\,dk = 1\;.
\end{equation}

The integration over the three-momentum in Eq.~(\ref{qq}) is
performed with the use of the delta function in the wave function
(\ref{vf}). The current matrix element that remains in the
integrand in the invariant variable $\sqrt{s}$ in (\ref{qq}) in
our approach is to be considered as a regular generalized function
determined on the space of test functions $\varphi(s)$, it is as a
distribution, i.e., as an object that makes sense only when it is
in the integrand. To parameterize it, we can proceed as when
obtaining Eqs.\ (\ref{breitsfer11}), (\ref{Atok11}), changing
$P_\mu$ for $P_{c\mu}$ in (\ref{lor}). So, we obtain the matrix
elements of the electroweak current for the system of two
interacting particles in the form analogous to (\ref{breitsfer11})
and (\ref{Atok11}):
$$
\frac{N_c}{N}\langle 0\,|j\,^c_0 (0)|\vec P_c\,,m_{c}\rangle =
0\;,
$$
$$
\frac{N_c}{N}\langle 0|j\,^{c1}_t(0)|\vec P_c,\sqrt{s},m\rangle =
$$
\begin{equation}
\langle 1\,m\, 1\, t |\,0\,0\rangle\,\frac{1}{\sqrt{4\pi}}\,
H^{1,0,1}_{0,1}(s)\;, \label{H}
\end{equation}
where $H^{1,0,1}_{0,1}(s)$ is a set of the reduced matrix elements, that
is  scalar functions, or form factors.

Substituting Eqs.\ (\ref{qq}), (\ref{vf}), (\ref{H}) into
Eq.~(\ref{const11}), we obtain the integral representation of the lepton
decay constant for a vector meson with zero orbital momentum of the quark
relative motion,
\begin{equation}
f_c =\int\,d\sqrt{s}\,H^{1,0,1}_{0,1}(s)\,\varphi(s)\;.
 \label{fc}
\end{equation}

In the frameworks of our method, the 4-fermion approximation is
formulated in the relativistic invariant way by using the
change of the invariant form factor
$H^{1,0,1}_{0,1}(s)$ in (\ref{fc})
for the invariant free form factor
$G^{1,0,1}_{0,1}(s)$, that enters the parametrization of the decay
current matrix element for the system of two free particles
(\ref{Atok11}):
\begin{equation}
f_c =
\int\,d\sqrt{s}\,G^{1,0,1}_{0,1}(s)\,\varphi(s)\;,\label{fc4-fi}
\end{equation}
with the free form factor $G^{1,0,1}_{0,1}(s)$ given by
(\ref{freeff}).

In the case of the $\rho $ meson,
one can rewrite Eq.(\ref{fc4-fi})
in a conventional form in terms of the variable
$k$,

\begin{eqnarray}\label{decayconst111}
&&\hspace{-2mm}f_{\rho}=\frac{\sqrt{3}}{\sqrt{2}
\pi}\int^{\infty}_{0}\,dk\,k^2\,\psi(k) \frac{(\sqrt{k^2 + M^2} +
M)}{(k^2 + M^2)^{3/4}}\nonumber\\
[2mm]&& \times\, \left(1 +
\frac{k^2}{3(\sqrt{k^2+M^2}+M)^2}\right)\;.
\end{eqnarray}

It is worth noting that Eq.\
(\ref{decayconst111})
coincides with analogous relations in other approaches (see, e.g.,
 \cite{Jau03, And11}, where the light-front dynamics and point form
dynamics were used).
However, some other expressions do differ, for example, our formulae  for
 electromagnetic form factors of composite systems
 (see, e.g., \cite{KrT03}).
This is caused by the fact that we introduce and use an alternative
variant of the relativistic impulse approximation (we called it the
modified impulse approximation, MIA).
In the MIA framework, contrary to the case of the standard impulse
approximation, the Lorentz covariance and the current conservation law
remain unbroken. In the non-relativistic limit, Eq.\
(\ref{decayconst111}) transforms to the standard form with the lepton
decay constant proportional to the wave function in coordinate
representation at the point $r=0$.

\section{The parameters and the results of calculations}
\label{sec:param}

For the calculation of the  $\rho$-meson characteristics basing on  the
relations (\ref{decayconst111})--(\ref{FF}) we use the following model
wave functions \cite{BrS95} (see, also
\cite{CoP05}).

1. A Gaussian or harmonic oscillator wave function
\begin{equation}
\psi(k)= N_{HO}\, \hbox{exp}\left(-{k^2}/{2\,b_\rho^2}\right).
\label{HOwf}
\end{equation}

2. A power-law wave function:
\begin{equation}
\psi(k) =N_{PL}\,{(k^2/b_\rho^2 + 1)^{-n}}\;,\quad n=2,3\;.
\label{PLwf}
\end{equation}

So, the following parameters enter our calculations:

1) the parameters that describe the constituent quarks {\it per
se} (the quark mass
$M$, the anomalous magnetic moments of quarks $\kappa_q$,
that enter our formulae through the sum
$s_q =
\kappa_u + \kappa_{\bar d}$, and the quark mean square radius (MSR) $\langle
r^2_q\rangle$);

2) the parameter  $b_\rho$ that enters the quark wave functions
(\ref{HOwf}),
(\ref{PLwf}) and is determined by the quark interaction potential.

In the paper \cite{KrT01} on pion we have shown that in our approach
all the parameters of the first group are the functions of the quark mass
$M$ and are defined by its value. In particular, for the quark MSR we can
use the relation (see, also \cite{CaG96}):
\begin{equation}
\langle r^2_q\rangle \simeq 0.3 /M^2\;.
\label{r2q}
\end{equation}

To calculate electroweak properties of
$\rho$-meson
we use the same values of quark parameters from the first group as that
we have used for the pion
\cite{KrT01}.
So, the wave function parameter
$b_\rho$ is
the only free parameter in our calculations.

Let us note that in papers \cite{CaG95,ChC04}, for calculations of
$\rho $-meson properties, quark wave functions of pion, i.e.\ $b_{\rho
}=b_{\pi }$, were used. It seems to us ineligible because it means that the
spin-spin interaction of quarks, which is different in $\rho $- and $\pi
$- mesons, is neglected.

Let us describe the procedure of calculation in detail, starting
from the quark parameter, $M=0.22$~GeV, used in a successful
calculation of the pion parameters \cite{KrT01}. As it has been
demonstrated in Ref. \cite{KrT01} (see also \cite{KrP15arxiv}),
the actual choice of the wave-function form does not affect the
result provided the quark parameters are fixed. In what follows,
we illustrate the procedure with the wave function (\ref{PLwf})
with $n=3$.

\begin{figure}
\begin{center}
\includegraphics[width=0.9\columnwidth]{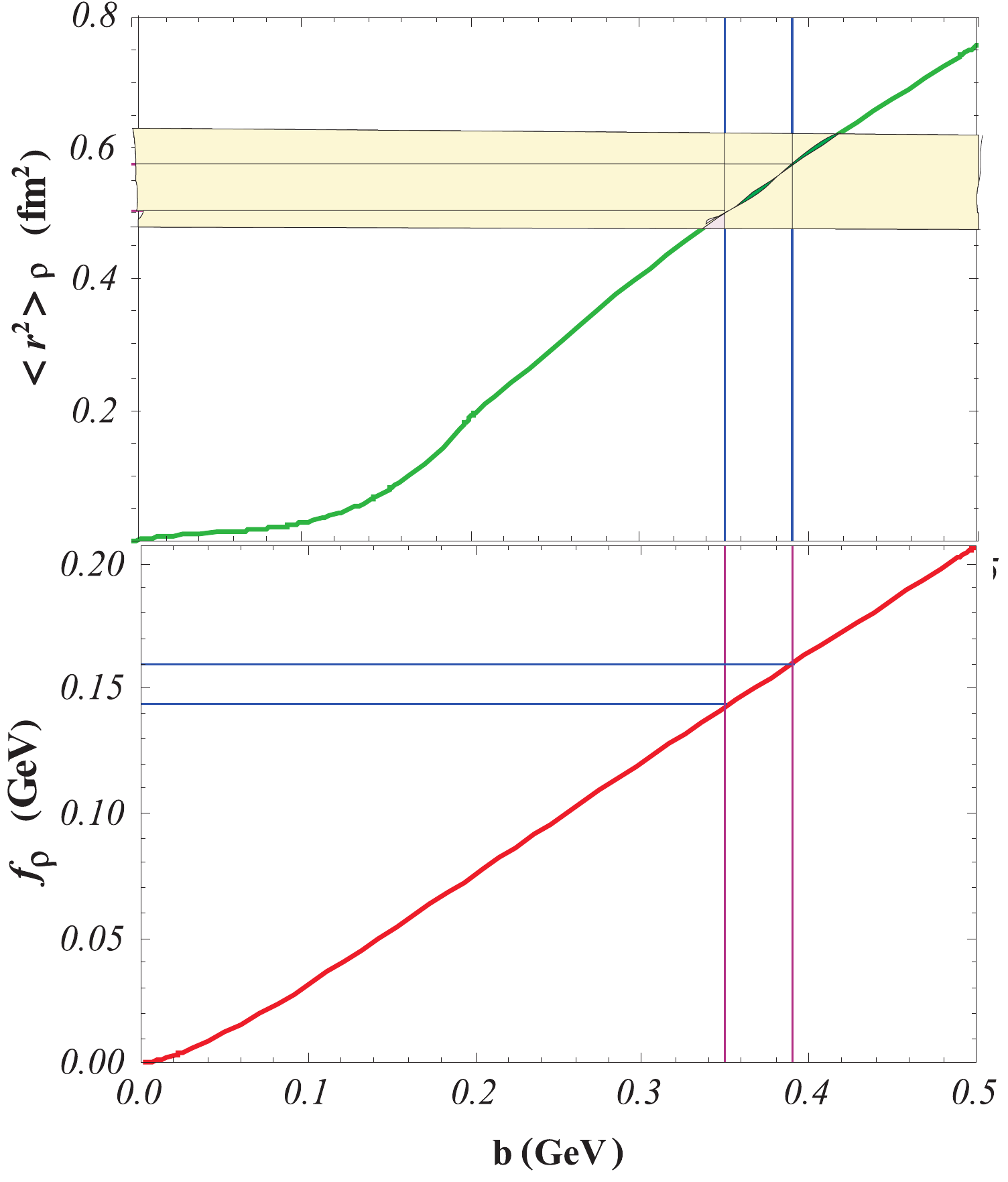}
\end{center}
\caption{ \label{constanta}
The decay constant
$f_{\rho}$ and the $\rho$-meson MSR, $\langle r^2_q\rangle$,  as functions
of the only free parameter of the model, $b_\rho$. Experimental data fix
the value of $f_{\rho}$, as shown in the lower panel. This fixes the value
of $b_{\rho}$, which determines in turn the value of  $\langle r^2_q\rangle$, as
shown in the upper panel. The values of quark parameters are taken from
Ref.~\cite{KrT01}; the wave function (\ref{PLwf}) with $n=3$ is used. }
\end{figure}

In the lower panel of Fig.~\ref{constanta},
the $\rho$-meson lepton decay constant $f_\rho$ as a function of
the only free parameter of the model, $b_\rho$, is presented. The
interval on the vertical axis representing the experimental values
of $f_\rho$, that is $f_{\rho}^{\rm exp} = (152 \pm 8)$~MeV
\cite{MeS15,Oli14}, is shown. It corresponds to the interval of
the values of $b_\rho$ which give, through our calculation, the
correct experimental values of the decay constant. This interval,
$b_\rho = (0.385 \pm 0.019)$~GeV, is shown on the horizontal axis
of Fig.~\ref{constanta}.

We turn now to the calculation of the $\rho$-meson charge radius for these
values of $b_\rho$.
We use the standard formula,
\begin{equation}\label{radius1}
\langle r^2_{\rho} \rangle =
-\left.6\frac{dG_{C}(Q^2)}{dQ^2}\right|_{\,Q^{2}\rightarrow 0}\;,
\end{equation}
where $G_{C}(Q^2)$ is the charge form factor of $\rho$-meson that was
obtained in \cite{KrT02} in the form
\begin{equation}\label{FF}
G_C(Q^2) = \int\,d\sqrt{s}\,d\sqrt{s'}\,
\varphi(s)\,g_{0C}(s\,,Q^2\,,s')\, \varphi(s')\;.
\end{equation}
Here, $g_{0C}$ is the free charge form factor that describes the
electromagnetic properties of non-interacting two particle composite system
with the quantum numbers of the $\rho $ meson,
\begin{eqnarray}
&&\hspace{-2mm}g_{0C}(s, Q^2, s') = A_{1}(s, Q^2, s')
(G^u_E(Q^2) + G^{\bar d}_E(Q^2))\,+\nonumber\\
[2mm]&& +A_{2}(s, Q^2, s')(G^u_M(Q^2) + G^{\bar d}_M(Q^2))\,\;,
\label{g0C}
\end{eqnarray}
where $G^{u,\bar d}_{E,M}$ are electric and magnetic form factors
of constituent $u$ and $\bar d$ quarks, respectively. The explicit
form of $A_{1}(s, Q^2, s')$ and $A_{2}(s, Q^2, s')$ is cumbersome;
it can be found in Ref.~\cite{KrP15arxiv2} which is an extended
version of Ref.~\cite{KrT02}.

The electromagnetic form factors of constituent quarks
are taken in the form
\cite{KrT09,KrT98,TrT13,TrT15},
$$
G^{q}_{E}(Q^2) = e_q\,f_q(Q^2)\;,
$$
\begin{equation}
G^{q}_{M}(Q^2) = (e_q + \kappa_q)\,f_q(Q^2)\;,
\label{q ff}
\end{equation}
where $e_q$ is the quark charge and $\kappa_q$ is the quark
anomalous magnetic moment,
\begin{equation}
f_q(Q^2) = \frac{1}{1 + \ln(1+ \langle r^2_q\rangle Q^2/6)}\; ,
\label{f_qour}
\end{equation}
$\langle r^2_q\rangle$ is the MSR of
the constituent quark. Values of all parameters used in these expressions
are taken from the $\pi$-meson calculation, see e.g.\ Ref.~\cite{TrT13}.

The calculated MSR of the $\rho$ meson is presented in the upper panel of
Fig.~\ref{constanta}. The interval of admissible values of $b_{\rho}$
gives now the corresponding interval of MSR predicted in the present study,
that our approach predicts: $ \langle r^{2}_\rho \rangle=(0.56 \pm
0.04)$~fm$^2$.

\section{Discussion}
\label{sec:discussion} Table~\ref{tab:4} presents a comparison of
our results with results of calculations of electroweak properties
of the $\rho $ meson  in other approaches.
\begin{table}
\caption{The lepton decay constant $f_{\rho}$ and MSR of the $\rho$ meson
calculated within different approaches.}
\label{tab:4}
\begin{tabular}{ccc}
\hline
Model  & $f_{\rho}$, MeV& $\langle r^2_{\rho} \rangle$, fm$^2$ \\
\hline
{\bf This work}  & 152$\pm$8  & 0.56$\pm 0.04$  \\
& (fixed) & \\
\hline
\cite{BhM08} & 146 & 0.54 \\
\hline
\cite{RoB11} & 130  & 0.312 \\
\hline
\cite{CaB15} & --- & 0.67 \\
\hline
\cite{LoM00} & ---  & 0.49 \\
\hline
\cite{ChA15} & 147.4 & --- \\
\hline
\cite{OwK15} & --- & 0.67 \\
\hline
\cite{GrR07} & ---  & 0.655\\
\hline
\cite{MeS02} & ---  & 0.33 \\
\hline
\cite{CaG95} & ---  &  0.35\\
\hline
\cite{BaC02} & 134  & 0.296 \\
\hline
\cite{ChC07} & 133  & --- \\
\hline
\cite{MeS15} & 154 & 0.268 \\
\hline
\end{tabular}
\end{table}

\begin{figure}
\begin{center}
\includegraphics[width=0.7\columnwidth]{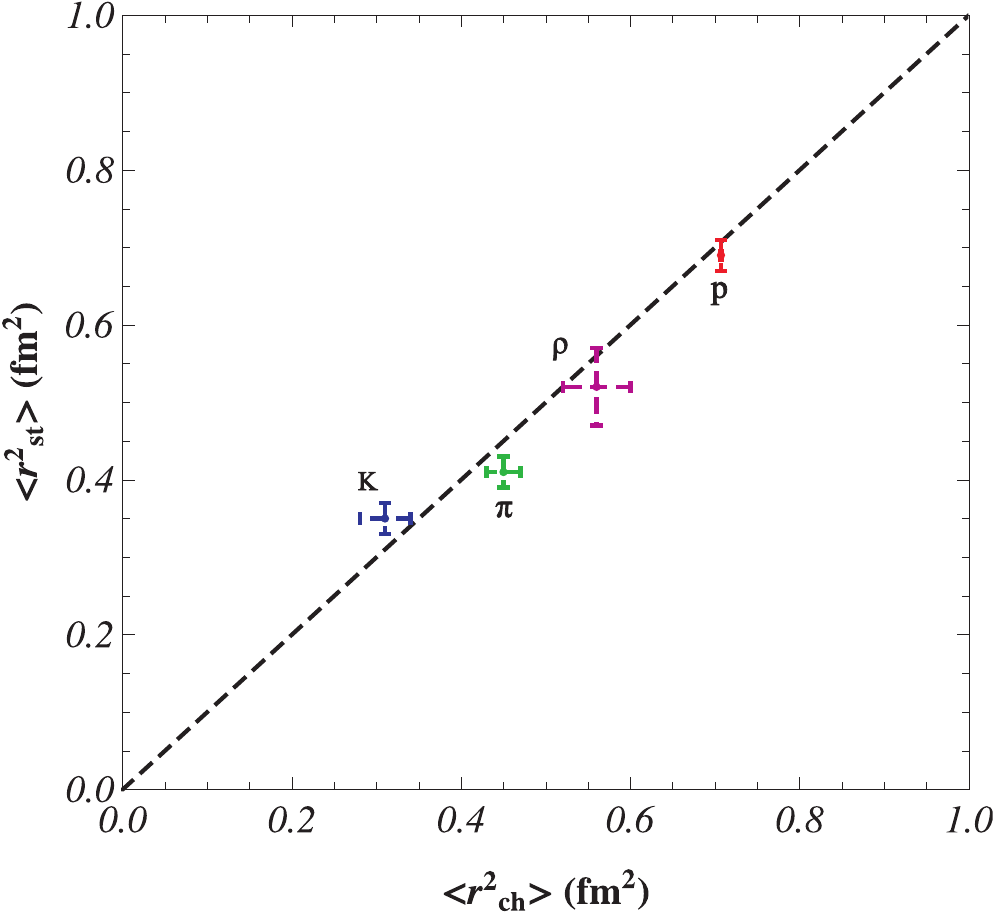}
\end{center}
\caption{ \label{radius} Relation between the strong-interaction
hadronic radius $\langle r^2_{\rm st}\rangle$ and the charge
radius $\langle r^2_{\rm ch}\rangle$ for light hadrons, see
Table~\ref{tab:1}.}
\end{figure}

The values of $\langle r^2_{\rho} \rangle$, while not measured
directly, are important for testing various conjectures about
strongly interacting systems. One of the interesting related
prediction was introduced as a consequence of the so-called
Wu--Yang hypothesis~\cite{WuY65} (see also Refs.~\cite{ChY68,
PoH87, PoH90, Gou74}), though it is remarkable by itself. Namely,
one may define the radius of a hadron either in terms of the
electroweak interaction (the mean square charge radius, $\langle
r^2_{\rm ch} \rangle$, calculated for the $\rho$ meson in this
paper) or in terms of the strong interaction (this radius,
$\langle r^2_{\rm st} \rangle$, is defined by the slope of the
cross section of hadron--proton scatering). The
conjecture~\cite{PoH90}, which may be derived from, though not
necessary implies, the hypothesis of Ref.~\cite{WuY65}, is the
equality of the two radii,
\begin{equation}
\langle r^2_{\rm st}\rangle = \langle r^2_{\rm ch}\rangle\;.
\label{rstch}
\end{equation}
This remarkable equality between two physical properties of a hadron
related to two different interactions of the Standard Model has been
verified experimentally with a great degree of accuracy for the proton,
$\pi$ and $K$ mesons (see Table~\ref{tab:1}).
\begin{table}
\caption{\label{tab:1}
The experimental values of the charge MSR $\langle r_{\rm ch}^2 \rangle$
and of the MSR for strong interaction $\langle r_{\rm st}^2 \rangle$.}
\begin{tabular}{ccc}
\hline Hadron & $\langle r_{\rm st}^2 \rangle$, fm$^2$ & $\langle
r_{\rm ch}^2
\rangle$, fm$^2$ \\
\hline
$\pi$  & 0.41$\pm$ 0.02 \cite{PoH90} & 0.45$\pm$ 0.02 \cite{Oli14}  \\
\hline
$K$  & 0.35$\pm$ 0.02 \cite{PoH90} & 0.31$\pm$ 0.03 \cite{Oli14} \\
\hline
$p$  & 0.69$\pm$ 0.02 \cite{Esc01} & 0.70706$\pm$0.00065 \cite{AnN13} \\
\hline
\hline
$\rho$  & 0.52$\pm$ 0.05 \cite{PoH90} & 0.56$\pm$0.04 ({\bf this work})\\
\hline
\end{tabular}
\end{table}
Even more demonstrative is Fig.~\ref{radius}, analogous to a figure from
the paper \cite{PoH90}, but presenting  more recent data.
We can see that the value of the $\rho$-meson charge radius obtained in
this paper fits perfectly the conjecture (\ref{rstch}).

\section{Conclusions}
\label{sec:concl} The present work gives a consistent unified
description of electroweak properties of both $\pi$ and $\rho$
mesons within a single relativistic-invariant approach and with a
single set of parameters of constituent quarks. Model is
constructed on the base of our variant of IF RQM containing the
modified impulse approximation (see, e.g., \cite{KrT02,KrT03}). In
the present paper, the expression for the lepton decay constant of
the $\rho$ meson, $f_\rho$, was derived in the frameworks of IF
RQM. For the derivation, a general method of the relativistic
invariant parametrization of local operators matrix elements
non-diagonal in the total angular momentum \cite{KrP15} was used.
Then, we turned to the calculation of the charge radius of the
$\rho$ meson, $\langle r_{\rho}^2 \rangle$. All but one parameters
of the model had been fixed by a successful description of the
$\pi$ meson in previous works. The remaining parameter was fixed
from the experimental value of $f_\rho$, thus allowing to derive
$\langle r_{\rho}^2 \rangle= (0.56 \pm 0.04)$~fm$^2$. This value
is in remarkable agreement with the strong-interaction radius,
$\langle r_{{\rm st,}\rho}^2 \rangle$ measured experimentally,
thus confirming the conjecture (\ref{rstch}) verified previously
for other light hadrons.

\begin{acknowledgments}
One of the authors (VT) thanks Sergey Troitsky for interesting
discussions and comments on the draft. This work was supported in
part (AK and RP) by the Ministry of Education and Science of the
Russian Federation (grant No. 1394, state task).
\end{acknowledgments}

\end{document}